\documentclass[a4paper,10pt,twoside]{article}
\usepackage{amsmath}
\usepackage{amssymb}
\usepackage{amsthm}
\usepackage{graphicx}

\theoremstyle{definition}
\theoremstyle{remark}

\newcommand{\bk}{{\bf k}}

\newcommand{\bR}{{\bf R}}
\newcommand{\bS}{{\bf S}}

\newcommand{\bB}{{\bf B}}

\newcommand{\bx}{{\bf x}}
\newcommand{\bz}{{\bf z}}
\newcommand{\by}{{\bf y}}
\newcommand{\bp}{{\bf p}}
\newcommand{\bC}{{\bf C}}
\newcommand{\bq}{{\bf q}}

\newcommand{\hk}{{\hat{\bf k}}}

\newcommand{\hx}{{\hat{\bf x}}}
\newcommand{\hy}{{\hat{\bf y}}}

\newcommand{\hp}{{\hat{\bf p}}}
\newcommand{\hq}{{\hat{\bf q}}}

\newcommand{\hz}{{\hat{\bf z}}}

\newcommand{\be}{\begin{equation}}
\newcommand{\ee}{\end{equation}}
\newcommand{\bay}{\begin{eqnarray}}
\newcommand{\eay}{\end{eqnarray}}
 \author{V.~S.~Buslaev, S.~B.~Levin}
 \title{The system of three three-dimensional charged quantum particles: asymptotic behavior of the eigenfunctions of the continuous spectrum at infinity}
\begin{document}
% \maketitle
 \addtolength{\hoffset}{-3.1cm}
  \addtolength{\voffset}{-3cm}

\maketitle

\begin{center}
{Department of Mathematical and Computational Physics,\\
St-Petersburg State University, Russia}
\end{center}

\begin{center}
PACS numbers: 03.65.Nk, 34.80.-i, 21.45.+v
\end{center}

\abstract{The asymptotic behavior in the leading order of the continuous spectrum eigenfunctions $\Psi(\bz,\bq)$
 as $|\bz|\rightarrow\infty$ for the system of
three three-dimensional charged quantum particles has been obtained on the heuristic level. The equality of the masses
and the equality of the absolute values of charges of particles are not crucial for the method.}

\vskip0.5cm

{\bf 1. Introduction.}
Consider the system of three three-dimensional quantum particles of equal masses interacting via  identical pair potentials. The original configuration space of the system is $\bR^9$. Stopping the motion of the centre of mass we come to a system on the configuration space  $\Gamma=\{\bz:\, \bz\in\bR^9,\,\bz=\{\bz_1,\bz_2,\bz_3\},\ \bz_1+\bz_2+\bz_3=0\}.$ On $\Gamma$ we consider the scalar product  $\langle\bz,\bz'\rangle$, induced by the standard  scalar product on  $\bR^9$. The system on  $\Gamma$ is described by the equation
$$
H\Psi=E\Psi,\ \ \ \Psi=\Psi(\bz)\in\bC,\ \ \bz\in\Gamma,\ \ \ \
H=-\Delta_\bz+v(\bx_1)+v(\bx_2)+v(\bx_3),\ \ \ \bx_j\in\bR^3.
$$
Here $\Delta_\bz$  is the Laplace operator  on $\Gamma$, $\ \bx_1=\frac{1}{\sqrt{2}}(\bz_3-\bz_2), \bx_2=\frac{1}{\sqrt{2}}(\bz_1-\bz_3), \bx_3=\frac{1}{\sqrt{2}}(\bz_2-\bz_1)$ . It is clear that  $\bx_1+\bx_2+\bx_3=0$.

Introduce also $\by_j=\sqrt{\frac{3}{2}}\bz_j$. It is not hard to check that on $\Gamma$ $\by_1+\by_2+\by_3=0,$
$$
\bz^2=\langle\bz,\bz\rangle=\langle\bx_j,\bx_j\rangle+\langle\by_j,\by_j\rangle,\ \ \ j=1,2,3\ \ \ , \quad \Delta_\bz =  \Delta_\bx +  \Delta_\by.
$$

Together with $\bz\in\Gamma,\ \bx,\by\in\bR^3$ we will consider the dual variables, momenta  $\bq \in \Gamma,\bk,\bp \in \bR^3$.

We will assume that
$$
v(\bx)=\frac{\alpha}{|\bx|},\ \ \alpha>0,
$$
although the generalization to the case
$
v(\bx)=\frac{\alpha}{|\bx|}+w(\bx),\ \ \ |\bx|w(\bx)\rightarrow 0,\ \ |\bx|\rightarrow\infty
$
is also possible.

{\bf 2.} Under the above assumptions the spectrum of the operator  $H$ is continuous and its eigenfunctions  $\Psi(\bz,\bq)$ can be numbered by the plane waves
$e^{i\langle\bz,\bq\rangle}$, that is  by the momenta  $\bq\in\Gamma$.  {\it It seems  that here we (obviously for the first time) obtain (although on the heuristic level) the asymptotic behavior in the leading order of the eigenfunctions   $\Psi(\bz,\bq)$ as $|\bz| \to \infty$, $\ \ \langle\bq,\hx_j\rangle\ge\varepsilon>0,\ \ j=1,2,3$.} We hope also
that the present work could be an essential step  on the way of the accurate proof of the results declared here.

If the potential $v(\bx)$ quickly decreased as  $|\bx|\rightarrow\infty$, the asymptotic behavior of the eigenfunctions
$\Psi(\bz,\bq)$ can be easily described by Faddeev's formula
\be
\Psi(\bz,\bq)\sim e^{i\langle\bz,\bq\rangle}+t(\bx_1,\bk_1)e^{i\langle\by_1,\bp_1\rangle}+
t(\bx_2,\bk_2)e^{i\langle\by_2,\bp_2\rangle}+
t(\bx_3,\bk_3)e^{i\langle\by_3,\bp_3\rangle},
\label{eq1}
\ee
\cite{LD1}.
Here $t(\bx,\bk)$ is characterized by the problem
$$
\psi(\bx,\bk)=(2\pi)^{-3/2}\left(e^{i\langle\bx,\bk\rangle}+t(\bx,\bk)\right),\ \ \ -\Delta_\bx\psi+v(\bx)\psi=k^2\psi, \quad
|\bk| = k,$$
where $t$ satisfies the radiation condition for $|\bx|\rightarrow\infty$. We don't estimate here the error of this formula, only remark that it correctly describes the leading term of the asymptotic behavior of $\Psi(\bz,\bq).$

Formula (\ref{eq1}) is not applicable in the case of the coulomb potentials.  However, as for the quickly decreasing
potentials,
the total coulomb potential $V$ after a natural simplification in the neighborhood of the screen  $\sigma_j=\{\bz:\,\bx_j=0\}$
allows the separation of variables.  The approximate solution constructed after the separation of variables outside of the neighborhood of  $\sigma_j$ has to be matched with the adiabatic modification of the plane wave
$e^{i\langle\bz,\bq\rangle}$. In the case the coulomb potentials this  adiabatic modification is known but is rather complicated, see
 \cite{BBK}. It is usually called
BBK--approximation.

In \cite{BL1} we have considered  the system of one-dimensional coulomb particles. The asymptotic behavior in this case essentially differs from the asymptotic behavior for the three-dimensional particles but there is something common in the logic of the constructing.

{\bf 3. BBK approximation.} For the coulomb potentials the standard radiation condition cannot be used for the description of the scattering of the plane waves $\psi_c(\bx,\bk),\ \bx,\bk\in\bR^3$ even for the system of two particles.  It is, however, known that this classical problem has the explicit solution:
$$
-\Delta_{\bx}\psi_c+\frac{\alpha}{|\bx|}\psi_c=k^2\psi_c,\ \ \ \ \ \psi_c(\bx,\bk)=N_c e^{i\langle\bx,\bk\rangle}D(\bx,\bk),
\ \ \ D(\bx,\bk)=\Phi(-i\eta,1,ikx-i\langle\bx,\bk\rangle),\ \ \ \eta=\frac{\alpha}{2k}.
$$
Here $\Phi$ -- the confluent hypergeometric function, see \cite{GR}, $N_c$ -- the normalization parameter
$N_c=(2\pi)^{-3/2}\Gamma(1+i\eta)e^{-\pi\eta/2}$.

In the system of the three coulomb particles in the adiabatic approximation the solution  $\Psi(\bz,\bq),\ \bz,\bq\in\Gamma,$ for $\bz\rightarrow\infty$
is described by the BBK formula, see  \cite{BBK},
$$
\Psi(\bz,\bq)\sim N_0 e^{i\langle\bz,\bq\rangle}D(\bx_1,\bk_1)D(\bx_2,\bk_2)D(\bx_3,\bk_3),\ \ \ \ \
N_0=N_c^{(1)}N_c^{(2)}N_c^{(3)},\ \ \ \ \ N_c^{(j)}=(2\pi)^{-3/2}\Gamma(1+i\eta_j)e^{-\pi\eta_j/2}.
$$
The discrepancy of this expression
$$
Q\left[\Psi(\bz,\bq)\right]=-\Delta_\bz\Psi+V(\bz)\Psi-E\Psi,\ \ \ \ V(\bz)=v(\bx_1)+v(\bx_2)+v(\bx_3),
$$
outside of the neighborhoods of  $\Omega_j,\ \ \ j=1,2,3,$ of the screens $\sigma_j$ decreases at infinity faster than the coulomb potential,
see  \cite{BBK}.

We will match this expression with the solution constructed with the help of the separation of variables.

{\bf 4. Weak asymptotic expansions. } Further instead of the ordinary (uniform) asymptotic formulas for the solution of the scattering problem we will often use the so called  {\it weak} asymptotic expansions. It will give us serious preferences.

The weak asymptotic expansions were introduced in  \cite{VSB} for the alternative description of the scattering matrix.
In \cite{MF} such asymptotic formulas were considered for the the system of two particles with the coulomb potential. Let us recall the main results of   \cite{VSB,MF}. The standard solution  $\psi(\bx,\bk)$ (of the standard plane wave type ) for the case of quickly decreasing potential
$$
-\Delta_\bx\psi+v(\bx)\psi=k^2\psi,\ \ \ \bx,\bk\in\bR^3,
$$
is characterized by the asymptotic behavior
$$
\psi(\bx,\bk)\sim e^{i\langle\bx,\bk\rangle}+f(\hx,\bk)\frac{e^{ikx}}{x},
$$
where $f$ -- is a smooth function of   $\hx$. This behavior can be also characterized as the asymptotic behavior
$x\rightarrow\infty$ in {\it the topology of the distributions } with respect to the variable $\hx\in\bS^2$:
\be
\psi(\bx,\bk)\sim\frac{2\pi i}{kx}\left(
\delta(\hx,-\hk)e^{-ikx}-S(\hx,\bk)e^{ikx}\right).
\label{q1}
\ee
Here $S(\hx,\bk)=\delta(\hx,\hk)-\frac{k}{2\pi i}f(\hx,\bk)$  is the two-particle scattering matrix , $f$ is the scattering amplitude.

In the case of the coulomb potential such asymptotic formula becomes more complicated. In particular, formula
(\ref{q1}) acquires the form
\be
\psi_c(\bx,\bk)\sim \frac{2\pi i}{kx}\left(\delta(\hx,-\hk)e^{-ikx+i\eta\ln x}-S_c(\hx,\bk)e^{ikx-i\eta\ln x}\right),
\label{q2}
\ee
where $S_c(\hx,\bk)=\frac{1}{2\pi}\frac{\Gamma(1+i\eta)}{|\hx-\hk|^{2+2i\eta}}2^{1+i\eta}e^{\frac{\pi\eta}{2}}$.

The notion of the weak asymptotic behavior can be used for the precise statement of the problem on the eigenfunction of the continuous spectrum for the system of three particles.
Let us define the solution  $\Psi(\bz,\bq)$ of the scattering problem for the system of three three-dimensional quantum particles with the help of the following weak asymptotic expansion for  $z\rightarrow\infty$:
\be
\Psi(\bz,\bq) \sim \frac12\left(\frac{4\pi i}{qz}\right)^{5/2}\delta(\hq,-\hz)e^{-iqz+\mathop{\sum}\limits_{j=1}^3\eta_j\ln z}-
\frac12\left(\frac{4\pi i}{qz}\right)^{5/2}S_c(\hq,\hz)e^{iqz-\mathop{\sum}\limits_{j=1}^3\eta_j\ln z}.
\label{coul}
\ee

In this definition $S_c$ remains  undefined and can be found in the course of searching of the solution. This coefficient is nothing but the scattering matrix for the corresponding three-particle problem.

We firmly believe in the following

{\bf Hypothesis 1.} {\it The solution $\Psi$ of the problem, defined by the condition (4), exists and is unique.}

Now we can formulate the goal of the work more precisely. We want to give the explicit description of the leading term of the asymptotic behavior of the solution   $\Psi$ in the uniform topology with respect to the angle  $\hz$. Namely this problem is solved here on the heuristic level. With this result we can change the statement of the problem, and search for the solution characterized by the uniform asymptotic behavior.

{\bf 5. BBK--approximation near the screen.} BBK-- approximation near, for example the screen  $\sigma_1$ naturally decomposes into the product
\be
\Psi_{BBK}(\bz,\bq)=\psi_c(\bx_1,\bk_1)\Psi_1(\bz,\bq),\ \ \ \Psi_1(\bz,\bq) = N_0^{(23)} e^{i\langle\by_1,\bp_1\rangle}
D(\bx_2,\bk_2)D(\bx_3,\bk_3),\ \ \ \ N_0^{(23)}=N_c^{(2)}N_c^{(3)}.
\label{bi0}
\ee
Near the screen the variables  $\bx_1,\ \by_1$ have asymptotically different orders, $y_1\gg x_1$. Let us compute the weak asymptotic expansion  $\Psi_1(\bz,\bq)$, as $y_1\rightarrow\infty$
\be
\Psi_1(\bz,\bq)=
\delta(\hat{\bp_1},-\hat{\by_1})B_0(\bq)\frac{2\pi}{iy_1p_1}e^{-iy_1p_1+i\omega\ln y_1}
e^{i\eta_2\ln\left[Z_2^- + \frac12\frac{x_1}{y_1}V_2^-\right]}
e^{i\eta_3\ln\left[Z_3^- + \frac12\frac{x_1}{y_1}V_3^-\right]}-
\label{bi11}
\ee
$$
-
\delta(\hat{\bp_1},\hat{\by_1})B_0(\bq)\frac{2\pi}{iy_1p_1}e^{iy_1p_1+i\omega\ln y_1}
e^{i\eta_2\ln\left[Z_2^+ + \frac12\frac{x_1}{y_1}V_2^+\right]}
e^{i\eta_3\ln\left[Z_3^+ + \frac12\frac{x_1}{y_1}V_3^+\right]}.
$$
We used here the following notations

$$
Z_2^\pm=\frac{\sqrt{3}}{2}(1\pm \langle\hat{\bp_1},\hat{\bk}_2\rangle),\ \ \ \ \ \
Z_3^\pm=\frac{\sqrt{3}}{2}(1\mp \langle\hat{\bp_1},\hat{\bk}_3\rangle),\ \ \ \ \ \
V_2^\pm=\langle\hx_1,\hk_2\pm\hp_1\rangle,\ \ \ \ \ \
V_3^\pm=\langle\hx_1,\hk_3\mp\hp_1\rangle,
$$
$$
\omega=\frac{\alpha}{2k_2}+\frac{\alpha}{2k_3},\ \ \ \ \ \ \ \
B_0(\bq)=-(2\pi)^{-3}k_2^{i\eta_2}k_3^{i\eta_3}.
$$
The coefficients $Z_{2(3)}^+,\ V_{2(3)}^+$ differ from the coefficients  $Z_{2(3)}^-,\ V_{2(3)}^-$
by the replacement of the vector  $\hp_1$ by the vector   $-\hp_1$.

{\bf 6. Separation of variables.}
Consider the whole potential $V(\bz)=v(\bx_1)+v(\bx_2)+v(\bx_3)$ in the vicinity of the screen $\sigma_1$. It is easy to see that  $\bx_2=-\frac{\sqrt{3}}{2}\by-\frac{1}{2}\bx,\ \
\bx_3=\frac{\sqrt{3}}{2}\by-\frac{1}{2}\bx,\ \ \bx=\bx_1,\ \ \by=\by_1$. We are interested in behavior of the solution at infinity so we have to assume $y\gg 1$. Therefore the formula
$$
V\sim v(\bx)+v_m(\by),\ \ \ v_m=\frac{4\alpha}{\sqrt{3} y},
$$
gives a good approximation to the potential. The equation with such potential allows the separation of variables
$$
-\Delta_\bz\chi + (v(\bx)+v_m(\by))\chi=E\chi.
$$

Since we are interested in the bounded solutions, for $\chi$ the following representation naturally appears
$$
\chi(\bx,\by,\bk,\bp)=\int\psi_c(\bx,\bk')\psi_m(\by,\bp')\delta({k'}^2+{p'}^2-E)R(\bq,\bq')d\bk' d\bp',\ \ \ \
\bq=(\bk,\bp),\ \ \bq'=(\bk',\bp').
$$
Here $\psi_c$ is the solution of the scattering problem for the potential  $v$, ànd $\psi_m$  is the solution of the scattering problem for the potential
$v_m$.

We will need the asymptotic behavior of  $\chi$ for $y\rightarrow\infty$. Substitute into the integral  $\chi$
the weak asymptotic representation of $\psi_m$:
\be
\chi\sim \int\psi_c(\bx,\bk)\frac{2\pi i}{p y}\left(\delta(\hy,-\hp)e^{-ipy+i\eta_m\ln y}-S_m(\hy,\bp)
e^{ipy-i\eta_m\ln y}\right)\delta({k'}^2+{p'}^2-E)R(\bq,\bq')d\bk' d\bp'.
\label{as1}
\ee
Here $S_m$ is the coulomb scattering matrix corresponding to the potential  $v_m$, i.e. to the parameter $\eta_m$.

{\bf 7. Computation of the coefficient  $R$.} For the further asymptotic simplification  of the integral  $\chi$ for $y\rightarrow\infty$ we need information on the structure of the coefficient
$R$. It can be obtained from the comparison of the integral with the  BBK-- approximation. This is a rather cumbersome computation that we are not able to demonstrate in this paper.
But, in fact, this computation is the most important part of the work.
The main idea of the computations is, however, quite natural. The representation of the $BBK-$ approximation  at some distance of the screen by the integral  $\chi$ is, in fact, the spectral resolution with the respect of the eigenfunctions  $\psi_m$ , and therefore, the resolution coefficient, i.e. the coefficient  $R$, can be found explicitly.

The result is:
\be
R(\bq,\bq')=\frac{1}{kpk'p'}A_{in}(\bq)\frac{\delta(\hat{\bp},\hat{\bp'})}{(p'-p+i0)^{1+ia}}
\delta\left(\hat{\bk}',\frac{\hat{\bk}+(p'-p)\bB_{in}}{|\hat{\bk}+(p'-p)\bB_{in}|}\right)+
\label{kern01}
\ee
$$
+\frac{1}{kpk'p'}A_{out}(\bq)\frac{G(\hat{\bp'},{\bp})}{(p'-p-i0)^{1+ib}}
\delta\left(\hat{\bk}',\frac{\hat{\bk}+(p'-p)\bB_{out}}{|\hat{\bk}+(p'-p)\bB_{out}|}\right).
$$
Here the kernel $G=S_m^{-1}$ satisfies the equation $\int S_m(\hy,p\hp')G(\hp',\bp)d\hp'=\delta(\hy,\hp)$,
$$
a=\omega-\frac{2\alpha}{\sqrt{3}p},\ \ \ \ \ b=\omega+\frac{2\alpha}{\sqrt{3}p},\ \ \ \ \
\omega=\frac{\alpha}{2k_2}+\frac{\alpha}{2k_3},
$$
The coefficient $A_{in}$ and the vector $\bB_{in}$ can be found from the equations
$$
A_{in}=-\frac{k}{\pi i}\Gamma(1-ia)e^{\frac{\pi a}{2}}B_0^{in},\ \ \ \ \ \ \
\bB_{in}=\frac{p}{k^2}\hk-\frac{1}{ak}\Omega_{in},\ \ \ \langle\bB_{in},\hk\rangle=0.
$$
Correspondingly, the coefficient  $A_{out}$ and the vector  $\bB_{out}$ can be found from the equations
$$
A_{out}=\frac{k}{\pi i}\Gamma(1-ib)e^{-\frac{\pi b}{2}}B_0^{out},\ \ \ \ \ \ \
\bB_{out}=\frac{p}{k^2}\hk-\frac{1}{bk}\Omega_{out},\ \ \ \langle\bB_{out},\hk\rangle=0.
$$
We used here the notations
$$
B_0^{in}(\bq)=(2\pi)^{-3}
\left[\frac{\sqrt{3}}{2}(1-\langle\hat{\bp},\hat{\bk}_2\rangle)\right]^{i\eta_2}
\left[\frac{\sqrt{3}}{2}(1+\langle\hat{\bp},\hat{\bk}_3\rangle)\right]^{i\eta_3}k_2^{i\eta_2}k_3^{i\eta_3},
$$
$$
B_0^{out}(\bq)=(2\pi)^{-3}
\left[\frac{\sqrt{3}}{2}(1+\langle\hat{\bp},\hat{\bk}_2\rangle)\right]^{i\eta_2}
\left[\frac{\sqrt{3}}{2}(1-\langle\hat{\bp},\hat{\bk}_3\rangle)\right]^{i\eta_3}k_2^{i\eta_2}k_3^{i\eta_3},
$$
$$
\Omega_{in}= \frac{1}{\sqrt{3}}\left(\eta_2\frac{\hat{\bk}_2-
\hat{\bp}}{1-\langle\hat{\bp},\hat{\bk}_2\rangle}+
\eta_3\frac{\hat{\bk}_3+
\hat{\bp}}{1+\langle\hat{\bp},\hat{\bk}_3\rangle}\right),\ \ \ \ \ \
\Omega_{out}= \frac{1}{\sqrt{3}}\left(\eta_2\frac{\hat{\bk}_2+
\hat{\bp}}{1+\langle\hat{\bp},\hat{\bk}_2\rangle}+
\eta_3\frac{\hat{\bk}_3-
\hat{\bp}}{1-\langle\hat{\bp},\hat{\bk}_3\rangle}\right).
$$

Note, this result for the simple case $v(\bx_1)=0$ corresponds to the result of the work \cite{AM}, for the case
$v(\bx_1)=\frac{\alpha}{|\bx_1|}$ we are not able to compare our results to the results of \cite{AM}.

With such a choice of  $R$ the asymptotic formula  (\ref{as1}) for $\chi$ and the asymptotic formula   (\ref{bi0})-(\ref{bi11})
for $\Psi^{BBK}_{j}$ as $y\rightarrow\infty$ coincide in the leading order on the intersection of the domains
$$
V_{0j}=\{\bz:\ x < y^{\nu}, 0 < \nu < 1\},\ \quad
V_{1j}=\{\bz:\ x > y^{\mu}, 1/2 < \mu < \nu\}.
$$

{\bf 8. Formulating of the result.} The sets  $V_{0j}$ and $V_{1j}$ cover $\Gamma$: $\Gamma =
V_{0j} \cup V_{1j}$. Consider the separation of the unit  $1 = \zeta _{0j}(x,y) + \zeta_{1j}(x,y)$, subordinated to this covering. Let us assume that on the subdomain where $\zeta _{0j}, \zeta _{1j}$ are not constant they depend on the ratio $\rho = lnx_j / lny_j$.

Consider the expression
$$
\Psi^{as}_j = \zeta_{0j} \chi_j.
$$
Let us define on $\Gamma$ the function
$$
\Psi^{as} = \sum_1^3 \Psi^{as}_j + (1 - \sum_1^3 \zeta_{0j})\Psi^{BBK}.
$$

We believe also in \\

{\bf Hypothesis 2.}
{\it The function $\Psi^{as}$ correctly describes the asymptotic behavior of the solution  $\Psi$ in the leading order.}

We also can check that the discrepancy  $Q[\Psi^{as}] = - \Delta_{\bf{z}} \Psi^{as} + V \Psi^{as} - E \Psi^{as}$ decreases as $z \to \infty$ faster than the coulomb potential.

\end{document}